\documentclass[preprint,final,12pt]{elsarticle}
\usepackage{amssymb}
\usepackage{mathtools}
\usepackage{amsmath}
\usepackage{type1cm}
\usepackage{geometry}
\usepackage{color}
\usepackage{epsfig}
\usepackage{latexsym}
\usepackage{setspace}
\usepackage{hyperref}
\usepackage[all]{hypcap}
\usepackage{graphicx}
\usepackage{epstopdf}
\usepackage{float}
\usepackage{subfig}
\usepackage{lineno}

\journal{Astroparticle Physics Journal \date{\today}}

\begin{document}

\begin{frontmatter}

\title{ Search for EeV Protons of Galactic Origin }

\author[]{\small \par\noindent
R.U.~Abbasi$^{1}$,
M.~Abe$^{2}$,
T.~Abu-Zayyad$^{1}$,
M.~Allen$^{1}$,
R.~Azuma$^{3}$,
E.~Barcikowski$^{1}$,
J.W.~Belz$^{1}$,
D.R.~Bergman$^{1}$,
S.A.~Blake$^{1}$,
R.~Cady$^{1}$
% ...

\par\noindent
{\footnotesize\it
$^{1}$ High Energy Astrophysics Institute and Department of Physics and Astronomy, University of Utah, Salt Lake City, Utah, USA \\
$^{2}$ The Graduate School of Science and Engineering, Saitama University, Saitama, Saitama, Japan \\
$^{3}$ Graduate School of Science and Engineering, Tokyo Institute of Technology, Meguro, Tokyo, Japan \\
% ...
}
\par\noindent
}

\begin{abstract} 
Cosmic rays in the energy range $10^{18.0}$ - $10^{18.5}$ eV are
thought to have a light, probably protonic, composition.  To study
their origin one can search for anisotropy in their arrival
directions. Extragalactic cosmic rays should be isotropic, but
galactic cosmic rays of this type should be seen mostly along the
galactic plane, and there should be a shortage of events coming from
directions near the galactic anticenter.  This is due to the fact
that, under the influence of the galactic magnetic field, the
transition from ballistic to diffusive behavior is well advanced, and
this qualitative picture persists over the whole energy range.  Guided
by models of the galactic magnetic field that indicate that the
enhancement along the galactic plane should have a standard deviation
of about 20$^\circ$ in galactic latitude, and the deficit in the
galactic anticenter direction should have a standard deviation of
about 50$^\circ$ in galactic longitude, we use the data of the
Telescope Array surface detector in $10^{18.0}$ to $10^{18.5}$ eV
energy range to search for these effects.  The data are isotropic.
Neither an enhancement along the galactic plane nor a deficit in the
galactic anticenter direction is found.  Using these data we place an
upper limit on the fraction of EeV cosmic rays of galactic origin at
1.3\% at 95\% confidence level.
\end{abstract}


\end{frontmatter}

\section{Introduction}

In studying ultrahigh energy cosmic rays, a fluorescence detector can
observe the development profile of extensive air showers initiated by
primary cosmic rays.  A detector with pixel size of 1$^\circ$, can
measure the depth of shower maximum, $X_{\mathrm{max}}$, to an
accuracy of about 20 g/cm$^2$ \cite{ta:mdhb_xmax}.  This is sufficient
to determine, on a statistical basis, the composition of primary
cosmic rays.  Although in some energy ranges there is disagreement
among experiments \cite{ta:mdhb_xmax,auger:xmax,hires:xmax}, from
about $10^{18.0}$ to $10^{18.5}$ eV, all fluorescence measurements
indicate that the composition is light, and probably protonic.

If this is the case, it is interesting to ask: are the sources of these
cosmic ray protons galactic or extragalactic.  One way to answer this
question is to look for anisotropy in their arrival directions.  While
extragalactic protons should be isotropic, one would expect that the
arrival directions of EeV protons of galactic origin should be
concentrated near the galactic plane.  The reason is that the critical
energy ($E_{\mathrm{C}}$) of the galactic magnetic field - the energy
where the Larmor radius equals the coherence length of the turbulent
component (\cite{Beck:2000dc}) - is thought to be about 0.3 EeV
\cite{Haverkorn:2014jka}.  At energies below $E_{\mathrm{C}}$ one
expects to see an isotropic distribution, since the turbulent
component randomizes proton directions.  Conversely, protons of
$10^{18.0}$ - $10^{18.5}$ eV should spiral around the regular
component of the field, but their directions should not be randomized
by the turbulent component.

In addition, if galactic cosmic ray sources were concentrated in the
disk, and there were many more sources at smaller galactic radii than
the galactic radius of the sun, there would be a second anisotropy
signal, a relative shortage of events in the direction of the galactic
anticenter.  For definiteness, in our model \cite{galprot_sim} we
chose the distribution of pulsars.  The details of this distribution
are not important, as long as the sources are concentrated in the disk
and their density is much larger than 1/kpc$^{3}$ \cite{galprot_sim}.
For a lower density of sources, our simulations show that an
unmistakable anisotropy would result from the sources nearest to the
Earth. Since we will see below that the data is isotropic to high
accuracy, this case is not realized in nature.

To search for these anisotropy signals, it would be useful to be able
to estimate the geometrical size of the effects: how wide should the
enhancement along the galactic plane be; and how wide should the
shortage of events be when looking along the galactic plane in the
direction of the galactic anticenter.  To perform these estimates, we
traced protons of $10^{18.0}$ - $10^{18.5}$ eV through the galactic
magnetic field (GMF) to the vicinity of the earth.  In galactic
coordinates ($l$,$b$), the result \cite{galprot_sim} is that the
enhancement in $b$ should be about 20$^\circ$ in standard deviation,
and the deficit centered at $l=140^\circ$ should have a standard
deviation of about 50$^\circ$.

These predictions are not sensitive to the details of the model
\cite{galprot_sim}.  For example, a large part of the enhancement
along the galactic plane comes from events whose sources are located
in the same galactic arm as the sun.  This type of enhancement will
occur for any model as long as the regular component of the GMF
follows the galactic arms.

In Section~\ref{section:gmf_pred} of this paper we describe the
properties of the GMF model and the prediction for anisotropy if
protons in the $10^{18.0}$ - $10^{18.5}$ eV range were of galactic
origin.  In Section~\ref{section:tasd} we describe the data collected
by the surface detector (SD) of the Telescope Array (TA) experiment.
In Section~\ref{section:results} we present our results and
conclusions.

\section{GMF Model and Anisotropy Prediction}
\label{section:gmf_pred}

The modeling of proton propagation in the GMF that we performed, prior
to searching for effects in the TA data, is described in reference
\cite{galprot_sim}.  In the galactic disk the magnetic field has two
components: a regular component which follows the galactic arms, and a
random component with critical energy 0.3 EeV.  In the energy range of
interest the effects of the random component are subdominant
\cite{Pshirkov:2013wka}.  Above and below the galactic disk the halo
component has a toroidal shape.  The main GMF model we use is that of
M. Pshirkov \emph{et~al.}  \cite{Pshirkov:gmf}.  The model of
R. Jansson and G. Farrar \cite{Jansson:gmf} gives very similar
quantitative answers.  Since the magnitude of the regular component of
the GMF at the location of the earth is slightly smaller in Jansson
and Farrar's model than in that of Pshirkov \emph{et~al.}, the
enhancement along $b$, and the deficit in $l$ around the galactic
anticenter are slightly narrower, and the effects we are searching for
are slightly larger.  The distribution of sources is assumed to be the
same as the distribution of pulsars in the galaxy given in reference
\cite{Lorimer:pulsars}.

It is clear that at energies $E < E_{\mathrm{C}}$ the random component
isotropizes the distribution of cosmic rays.  If there are many
sources, one would see a continuous distribution in ($l$,$b$).  Above
$E_{\mathrm{C}}$, for instance for $10^{18.0}$ - $10^{18.5}$ eV, the
random component of the GMF cannot isotropize cosmic rays.  If the
source density is high, a continuous distribution (no small-scale
excesses or deficits) will result.  If the source density is low, the
nearest sources will produce small-scale anisotropy signals, which are
not visible in the data described later in this paper.

\begin{figure*}[!ht]
  \centering  
  \subfloat[]{\includegraphics[width=0.5\textwidth]{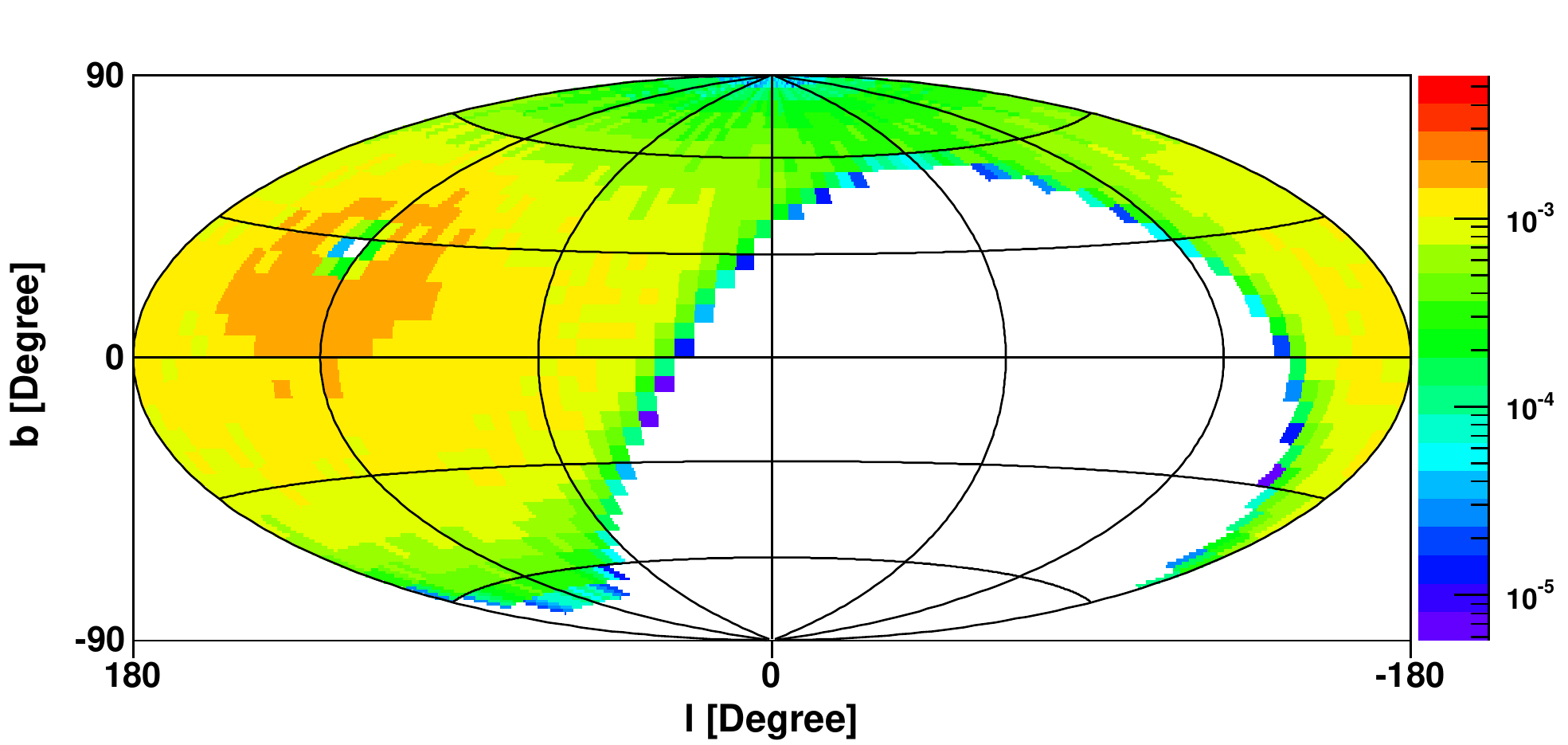}\label{figure:isopred}}
  \subfloat[]{\includegraphics[width=0.5\textwidth]{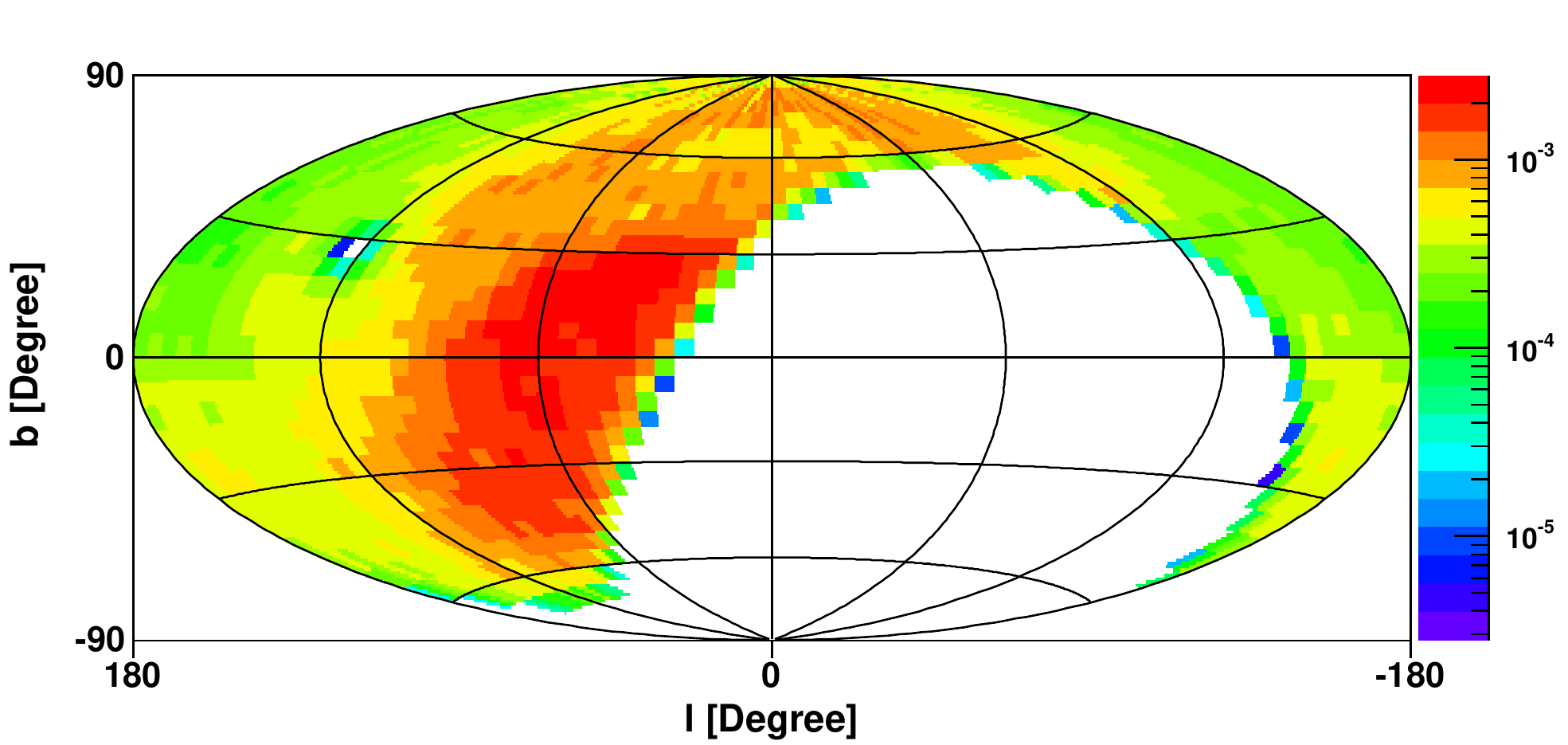}\label{figure:gmfpred1}}
  \linebreak
  \subfloat[]{\includegraphics[width=0.5\textwidth]{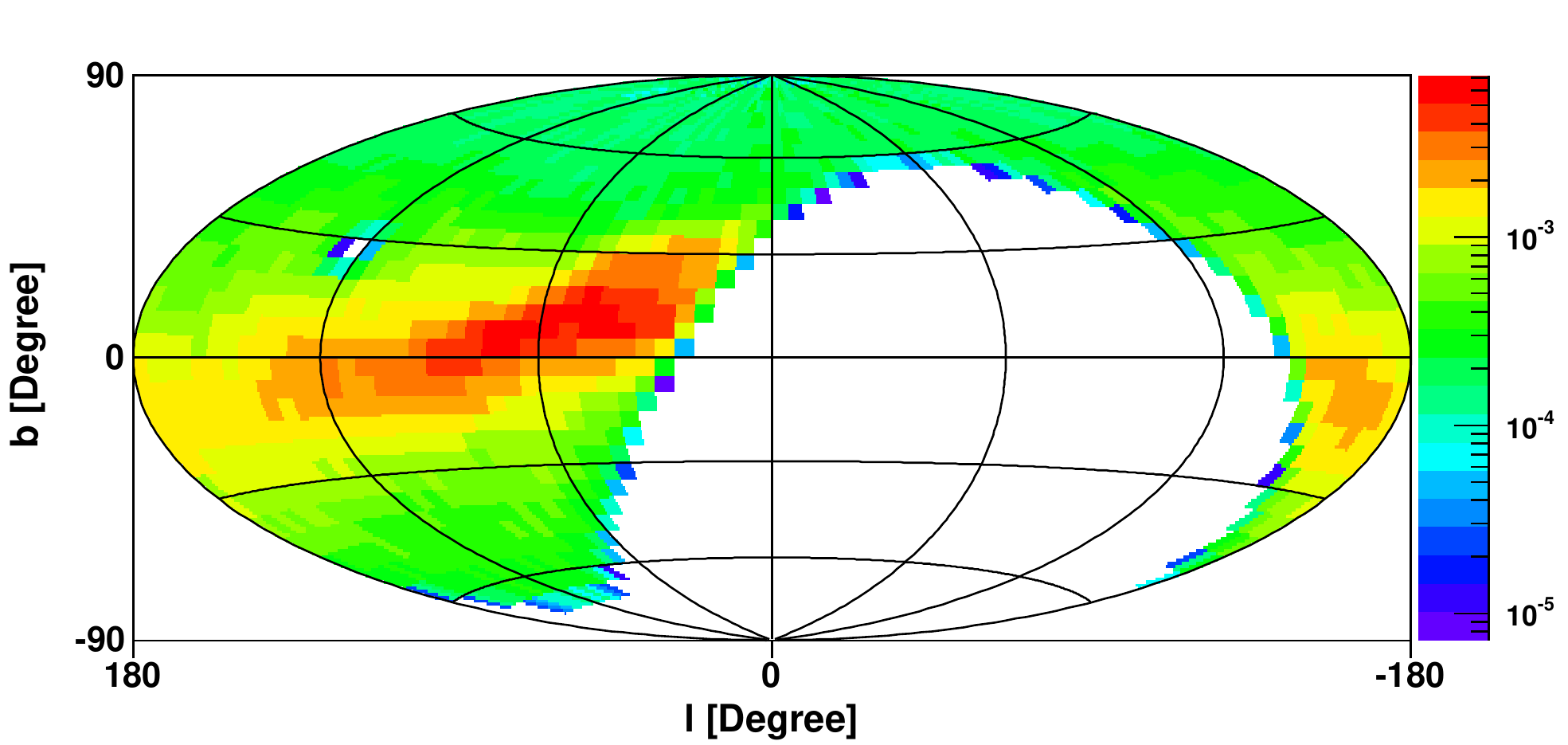}\label{figure:gmfpred2}}
  \subfloat[]{\includegraphics[width=0.5\textwidth]{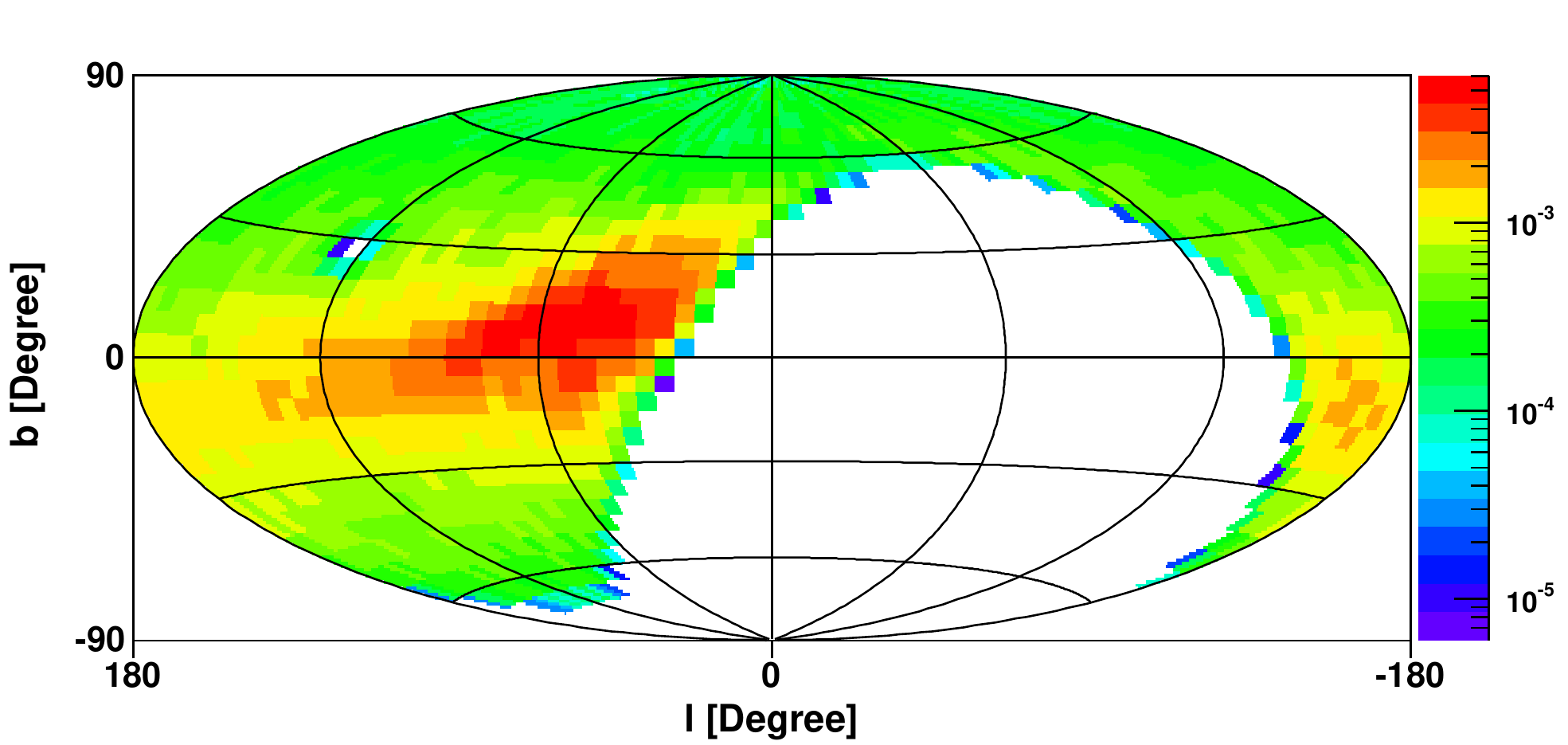}\label{figure:gmfpred3}}
  \caption{ Normalized flux distributions, represented by color, as
    would be seen by the TA SD experiment.
    \protect\subref{figure:isopred} In the case of isotropy;
    \protect\subref{figure:gmfpred1} and
    \protect\subref{figure:gmfpred2} In the case of 1 and 3 EeV
    galactic protons, respectively; \protect\subref{figure:gmfpred3}
    In the case of $10^{18.0} - 10^{18.5}$ eV galactic protons
    simulated using measured cosmic ray energy spectrum.
    \protect\subref{figure:gmfpred1} -
    \protect\subref{figure:gmfpred3} use Pshirkov \emph{et~al.} model of the
    GMF.  Anisotropies predicted by the models of galactic protons are
    evident in \protect\subref{figure:gmfpred1} -
    \protect\subref{figure:gmfpred3}, while
    \protect\subref{figure:isopred} shows the non-uniformities that
    are expected from the zenith angle dependence of the TA SD
    acceptance in the 10$^{18.0}$ to 10$^{18.5}$ eV range. The deficit
    at $l \sim 123^\circ$, $b \sim 27^\circ$ in all figures
    corresponds to the equatorial North Pole, which is not seen by the
    TA SD due to the maximum event zenith angle cut of 45$^\circ$ in
    this analysis. }
  \label{figure:pred}
\end{figure*}

Figures~\ref{figure:gmfpred1} - \ref{figure:gmfpred2} show sky maps of
the prediction of the Pshirkov \emph{et~al.} model for galactic
protons at 1 and 3 EeV, respectively.  Figure~\ref{figure:gmfpred3}
shows the model prediction for the galactic protons with energies
distributed according to the measured spectrum of cosmic rays in the
$10^{18.0} - 10^{18.5}$ eV range,
\cite{hires:spectrum,ta:sdspec_di_thesis,TheTelescopeArray:2015mgw,ThePierreAuger:2015rha,Verzi:2016ara,ta:spectrum_summary,auger:spectrum_summary},
where the High Resolution Fly's Eye, Telescope Array, and Pierre Auger
experiments are in good agreement, when their measurements are
adjusted to use a common energy scale.

Figure~\ref{figure:isopred} shows the sky map if
the distribution were isotropic.  The anisotropies show up clearly in
Figures~\ref{figure:gmfpred1} - \ref{figure:gmfpred3}.
\begin{figure*}[!ht]
  \centering
  \subfloat[]{\includegraphics[width=0.5\textwidth]{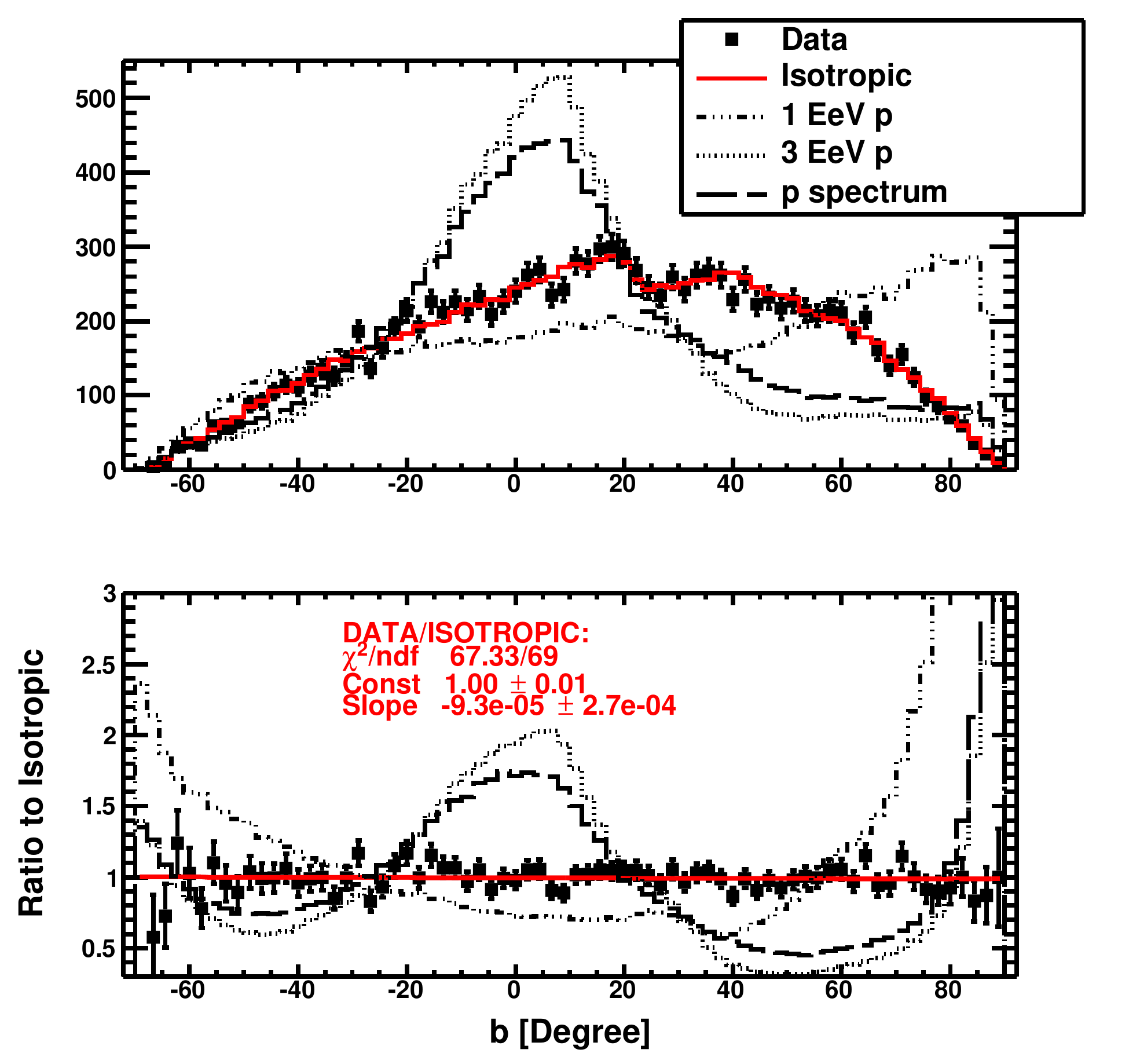}\label{figure:bcomp}}
  \subfloat[]{\includegraphics[width=0.5\textwidth]{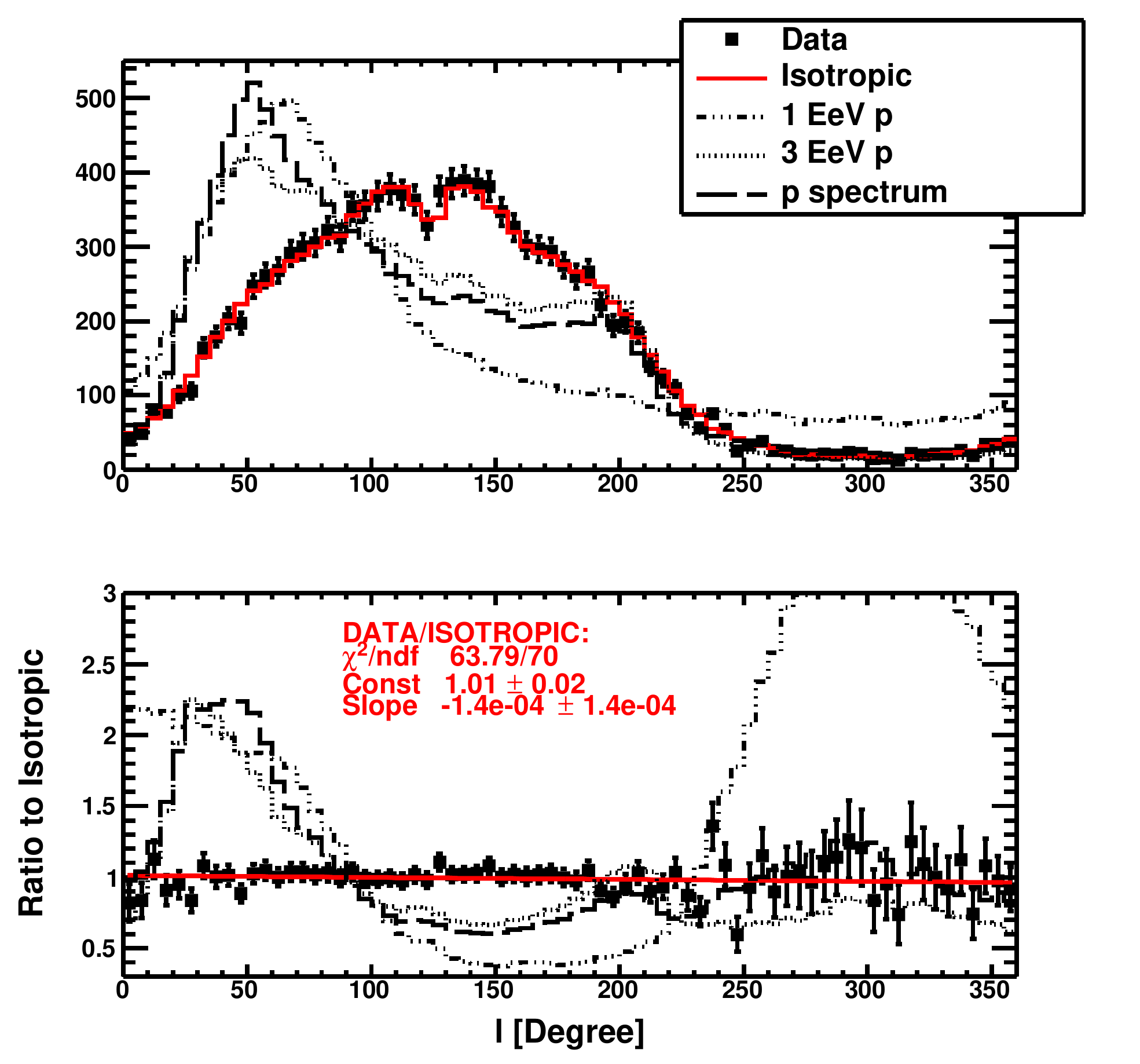}\label{figure:lcomp}}
  \caption{ Histograms of galactic latitude, $b$ and galactic
    longitude, $l$, made from the sky maps of galactic protons of 1
    EeV (dashed-dotted line), 3 EeV (dotted line), and galactic
    protons following measured cosmic ray energy spectrum in the
    $10^{18.0}$ - $10^{18.5}$ eV range (dashed line).  The histograms
    of the mono-energetic galactic proton models, as well as the
    histogram of the isotropic Monte-Carlo (solid red line) have been
    normalized to the total number of TA SD data events (black points
    with error bars), and the ratios of all model and data histograms
    to the isotropic distribution are shown in the bottom parts of the
    figures.  Linear fits are made to the ratios of the data
    histograms to the isotropic histograms in $l$ and $b$ (red lines
    in the bottom figures), and the result (shown in red letters on
    the plots) is that the values of the slopes are within their
    fitting uncertainties and the reduced $\chi^{2}$ values of the
    fits are close to unity.  The data is consistent with isotropy
    (see also \cite{Auger:2012an,Abreu:2012ybu}) but inconsistent with
    any of the galactic proton models. }
  \label{figure:blcomp}
\end{figure*}
Figure~\ref{figure:bcomp} shows a histogram of events in $b$ as it
would be seen by the TA experiment, and Figure~\ref{figure:lcomp}
shows the histogram in $l$.  This figure is a prediction of what would
be seen if all cosmic rays in the $10^{18.0}$ - $10^{18.5}$ eV energy
range were protons of galactic origin.  Also shown are the $b$ and $l$
of the TA data, and histograms of an isotropic distribution, according
to the TA surface detector Monte Carlo simulation.  All histograms are
normalized to the same area.  The galactic and isotropic distributions
are clearly different, and a measurement of the $l$ and $b$
distributions of cosmic rays in the $10^{18.0}$ - $10^{18.5}$ eV
energy range should allow a determination of the fraction of events of
galactic origin.  A hint of the result is already shown in the
Figure~\ref{figure:blcomp}.  It is worth emphasizing that the
prediction of the isotropic Monte Carlo simulation is robust (the TA
Monte Carlo simulation is very accurate), but there is some model
dependence in the exact shape of the $l$ and $b$ distributions from
galactic sources.  If the data were to resemble the galactic model, or
if there were a mixture of galactic and extragalactic sources, one
would perform a fit to the sum of isotropic and galactic models to
determine the ratio of events from the two types of sources.  One
might have to vary the galactic model shape to get a good fit.  If the
data were to resemble the isotropic simulation then one would search
for a deviation from that simulation by using the widths of the
expected galactic distributions.  In Section~\ref{section:results} we
will see that there is no ambiguity in the result.

\section{Telescope Array Surface Detector Data}
\label{section:tasd}

To perform the search we used the data of the surface detector of the
Telescope Array experiment.  TA is the largest experiment studying
ultrahigh energy cosmic rays in the northern hemisphere.  The TA
experiment consists of a surface detector (SD) \cite{ta:sd} covering
~700 km$^{2}$, and 48 fluorescence telescopes
\cite{ta:brlrfd,ta:mdfd}, located at three sites, which overlook the
SD array.  The data described here were collected over 7 years, from
11 May, 2008 to 11 May, 2015.

The TA SD consists of 507 scintillation counters each 3 m$^{2}$ in
area, deployed in a grid of 1.2 km spacing in the desert of Millard
County, Utah, USA.  There are 2 layers of plastic scintillator in each
counter.  The counters are solar-powered, read out by a radio system,
and calibrated using single muons every 10 minutes, to determine the
pulse height of a minimum ionizing particle (MIP).  The SD trigger is
satisfied when within an 8$\mathrm{\mu}$s window 3 adjacent counters
have energy deposits equivalent to 3 MIPs or more.  Every second the
counters are queried as to the times of 3 MIP hits, and when the
trigger conditions are satisfied, all counters are instructed to
report the FADC wave forms of hits above 0.3 MIP energy deposit.

TA SD data analysis consists of two steps: the first is a fit to the
time that counters are struck, to determine the direction of the air
shower, and the second is a fit to the distribution of counter pulse
heights as a function of distance perpendicular to the path of the air
shower.  The signal at a distance of 800 m from the shower core,
called S800, is determined and the cosmic ray energy is found from a
look-up table with inputs S800 and sec($\theta$), where $\theta$ is
the zenith angle of the shower.  The energies come from a Monte Carlo
simulation of the SD using CORSIKA \cite{corsika} and the hadronic
generator QGSJET II-3 \cite{qgsjet}.  The TA SD energy resolution is
estimated from the Monte Carlo to be 36\% for $10^{18.0}\mathrm{eV} <
E < 10^{18.5}\mathrm{eV}$, 29\% for $10^{18.5}\mathrm{eV} < E <
10^{19.0}\mathrm{eV}$, and 19\% for $E > 10^{19.0}\mathrm{eV}$
\cite{ta:sdspec_di_thesis}.  The TA SD angular resolution is 2.4$^{o}$
for $10^{18.0}\mathrm{eV} < E < 10^{18.5}\mathrm{eV}$, 2.1$^{o}$ for
$10^{18.5}\mathrm{eV} < E < 10^{19.0}\mathrm{eV}$, and 1.4$^{o}$ for
$E > 10^{19.0}\mathrm{eV}$ \cite{ta:sdspec_di_thesis}.

Although the approximation technique called thinning is used in our
Monte Carlo simulation, we carry out a process called de-thinning
\cite{dethinned} to restore lost shower information.  Upon comparing
the energies of cosmic rays found in this way, and those found for the
same events using the data of our fluorescence detectors, we see a
difference of a constant fraction of 1.27 \cite{ta:sdspec_5yr},
independent of energy.  Because the fluorescence detector energy
measurement has a smaller systematic uncertainty, since it is a
calorimetric determination, we correct the SD energy (downward) by
this factor.

\begin{figure}[!ht]
  \begin{center}
    \includegraphics[width=0.5\textwidth]{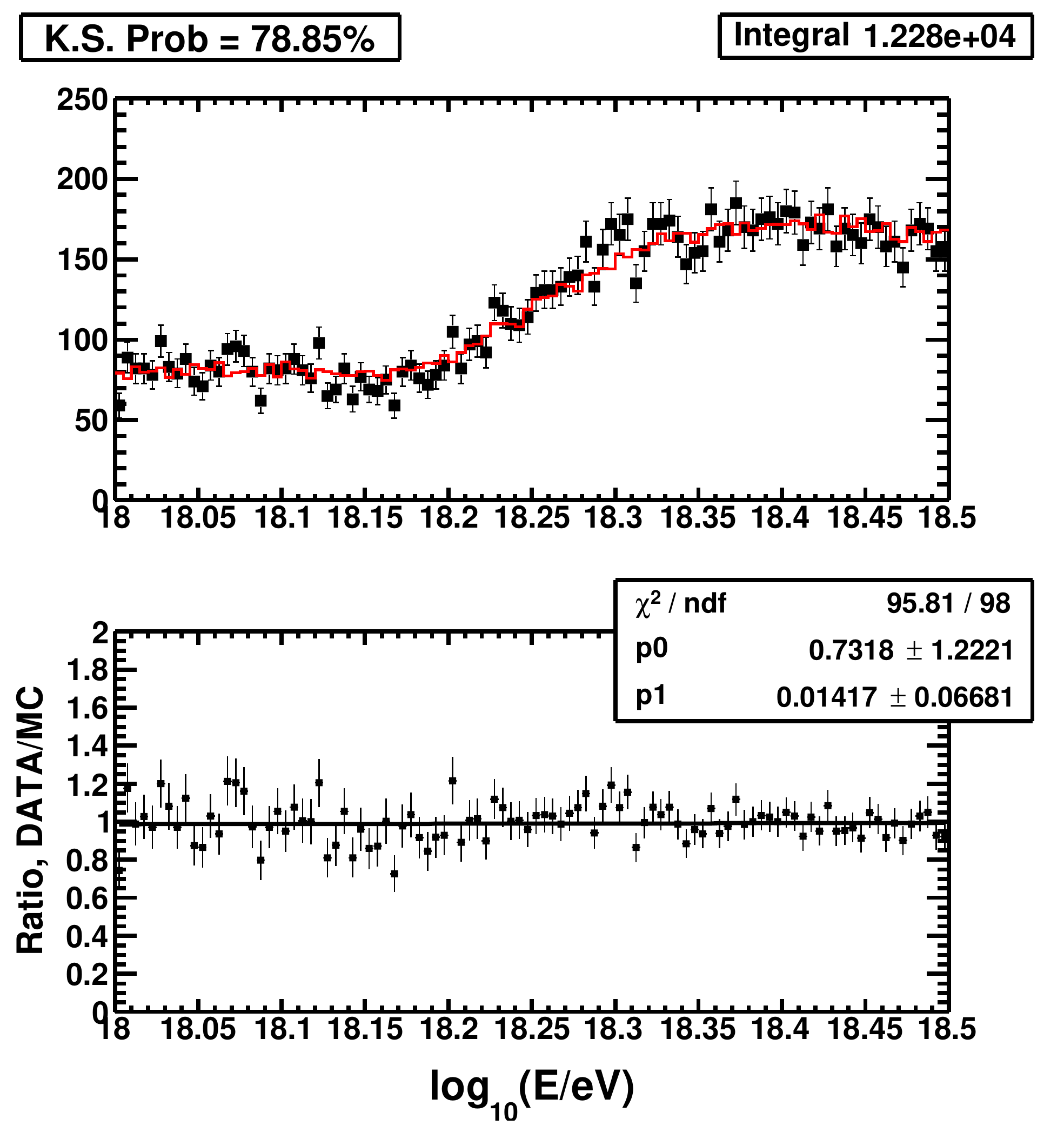}
  \end{center}
  \caption{ Histogram of TA surface detector data in log$_{10}$ of
    energy in eV units. Black points show the data and the red line is
    the Monte-Carlo histogram. When the data and Monte-Carlo
    histograms are compared, the Kolmogorov-Smirnov probability is
    79\% (top) and the linear fit of the ratio of the data to
    Monte-Carlo distributions (bottom) has a slope of $0.014 \pm
    0.067$, indicating a good agreement.  The mean event energy is
    2.06 EeV. }
  \label{figure:dtmc_energy}
\end{figure}

\begin{figure*}[!ht]
  \centering
  \subfloat[]{\includegraphics[width=0.5\textwidth]{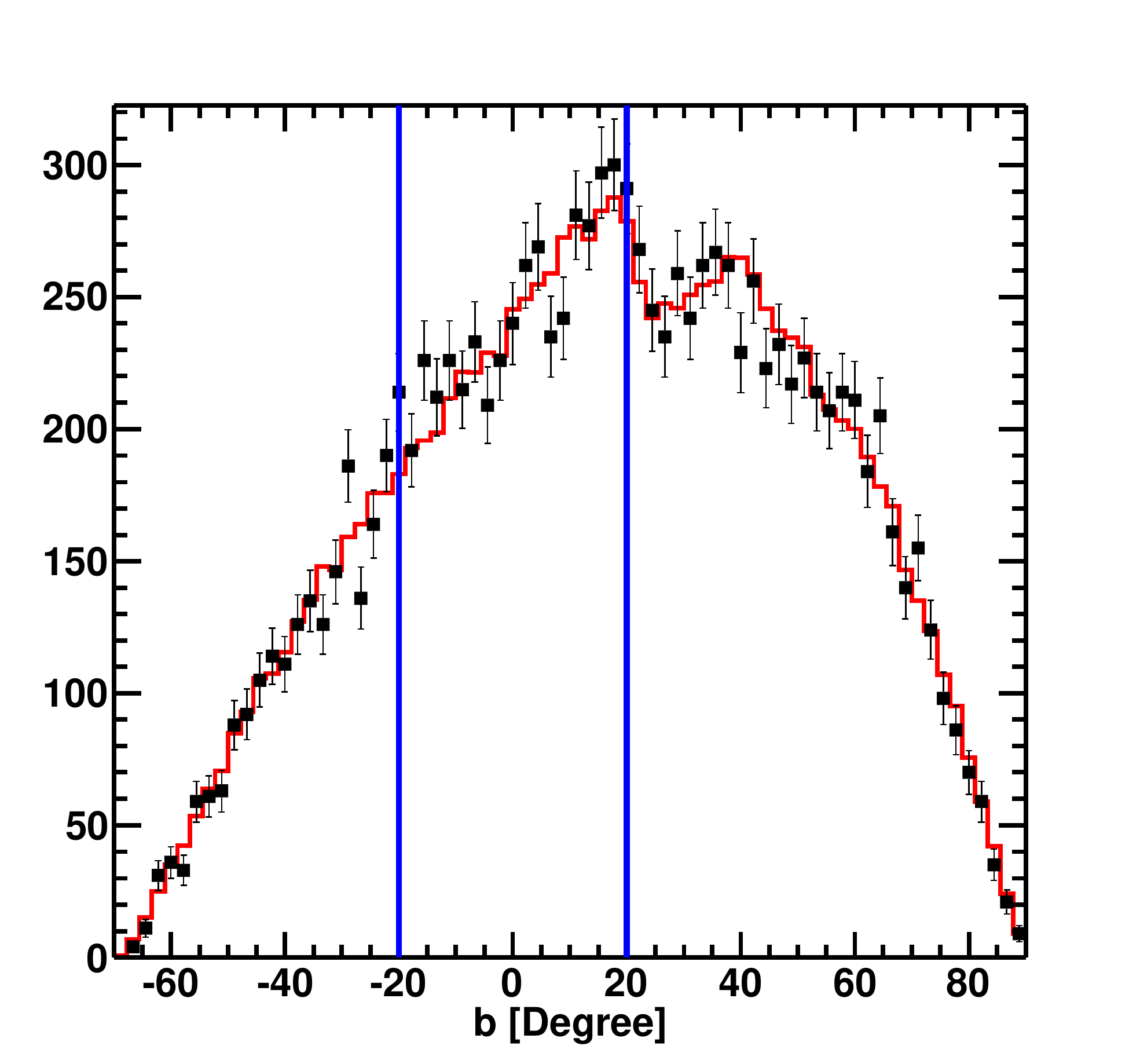}\label{figure:bhist}}
  \subfloat[]{\includegraphics[width=0.5\textwidth]{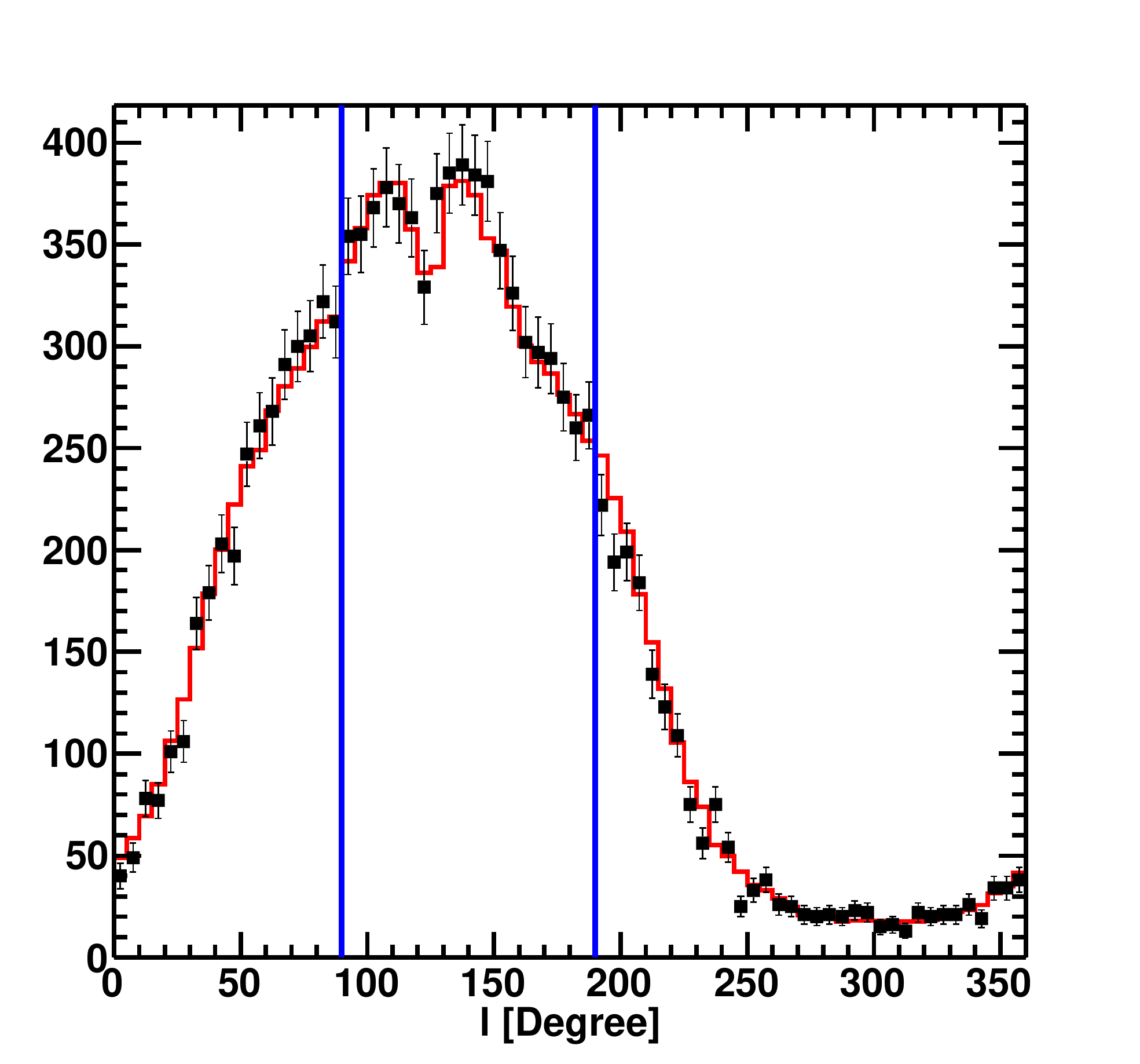}\label{figure:lhist}}
  \caption{Histogram of TA surface detector data in galactic latitude,
    $b$ (left panel) and galactic longitude, $l$ (right panel).  TA
    data are the black points with error bars, and the TA Monte Carlo
    simulation is the red histogram.  The blue lines, at $\pm
    20^\circ$ in the left panel, and at $\pm$50$^\circ$ in the right
    panel, show where the enhancement in $b$ and deficit in $l$ are
    expected for galactic protons.  Neither effect is seen in these
    data.  The upper limits on the galactic proton flux (95\%
    Confidence Level) from examining these histograms are 1.3\% when
    we search in $b$, and 0.3\% if we search in $l$.  }
  \label{figure:blhist}
\end{figure*}
Choosing events of energies between $10^{18.0}$ and $10^{18.5}$ eV,
(the energy histogram is shown in Figure~\ref{figure:dtmc_energy}), we
show in Figure~\ref{figure:blhist} histograms of the galactic
latitude, $b$, and longitude, $l$.  The data is shown as black points
with error bars, and the red histogram is the prediction of our Monte
Carlo simulation for an isotropic distribution.  The blue lines
indicate the range over which an excess in $b$, and a deficit in $l$,
are expected if $10^{18.0}$ - $10^{18.5}$ eV cosmic rays come from
galactic sources.  No excess at low galactic latitudes is evident, nor
does a deficit of events show up at $l$ = 140$^\circ$, the direction
away from the galactic center.

\section{Results}
\label{section:results}

We search for evidence for galactic origin of cosmic rays in our data
using two approaches.  In the first approach, we compare the galactic
latitude and longitude histograms shown in Figure~\ref{figure:blhist}
to the Monte Carlo distributions of isotropic events. This method is
less sensitive to the details of the model of the galactic protons:
the only model-dependent information used in this approach is that
there is an expected excess for $|b| < 20^\circ$ and a deficit for
$|l-140^\circ|<50^\circ$, if we simulate galactic protons using
measured cosmic ray spectrum in $10^{18.0} - 10^{18.5}$ eV range.

In Figure~\ref{figure:blhist}, the Monte Carlo is normalized to the
total number of events in the data.  We calculate the fractional
difference, $F = (N_{\mathrm{Data}} -
N_{\mathrm{MC}})/N_{\mathrm{Total}}$ between the data and Monte Carlo
while applying a cut on galactic latitude, $b$: $|b| < 20^\circ$.  We
then calculate $F$ while applying a cut $|l-140^\circ| < 50^\circ$ in
$l$. Here $N_{\mathrm{Data}}$ and $N_{\mathrm{MC}}$ are the number of
data and isotropic Monte Carlo events that pass the cuts on $b$, and
$l$ separately, and $N_{\mathrm{Total}}$ is the total number of events
in our data between $10^{18.0}$ and $10^{18.5}$ eV, which is 12281 for
the 7 years of TA SD data.  No significant enhancement in $b$ or
deficit in $l$ was found.  The 95\% confidence level upper limit on
$F$ from this study is found to be 1.3\% for the $b$ analysis, and
0.3\% for the $l$ analysis.

\begin{figure*}[!ht]
  \centering
  \subfloat[]{\includegraphics[width=0.5\textwidth]{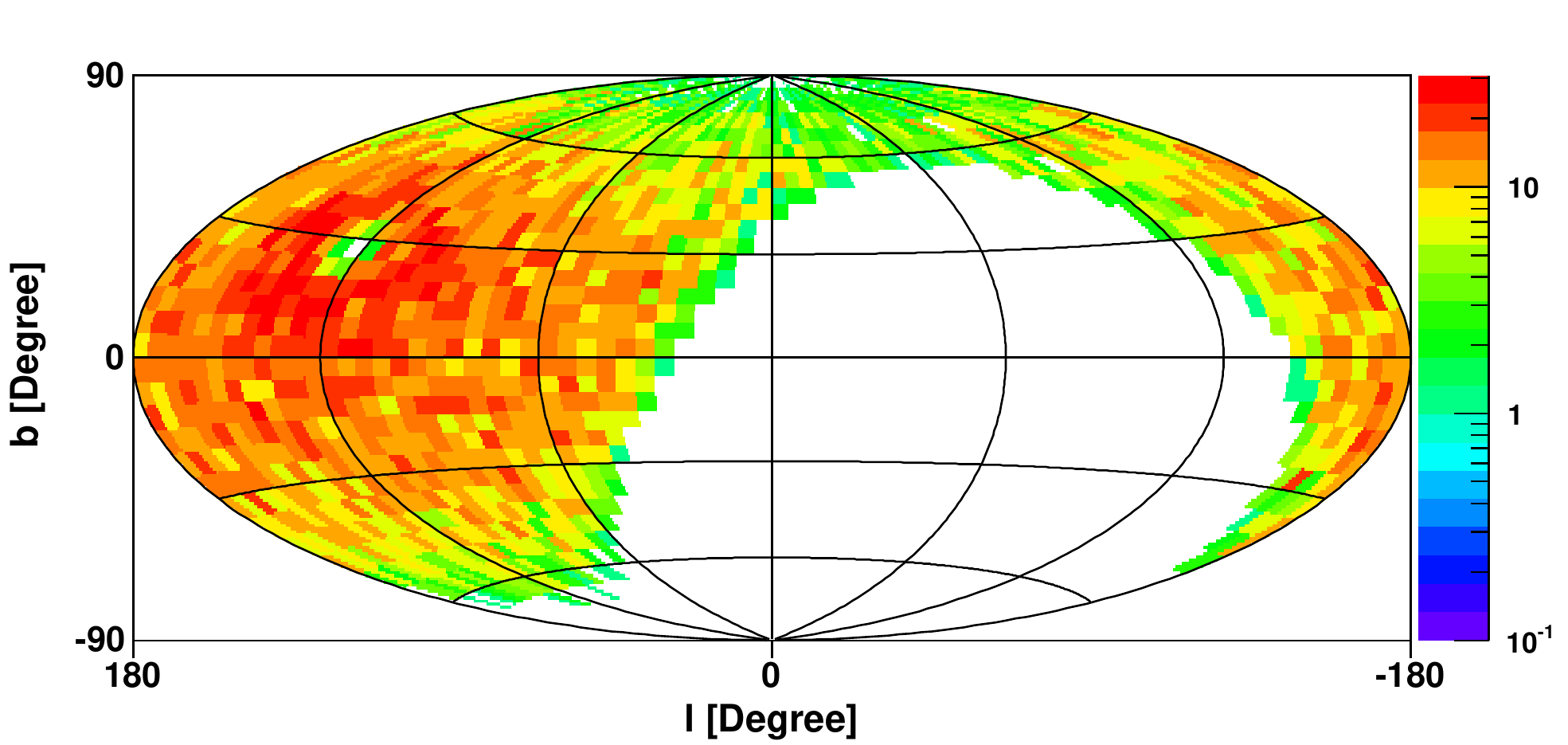}\label{figure:data2d}}
  \subfloat[]{\includegraphics[width=0.5\textwidth]{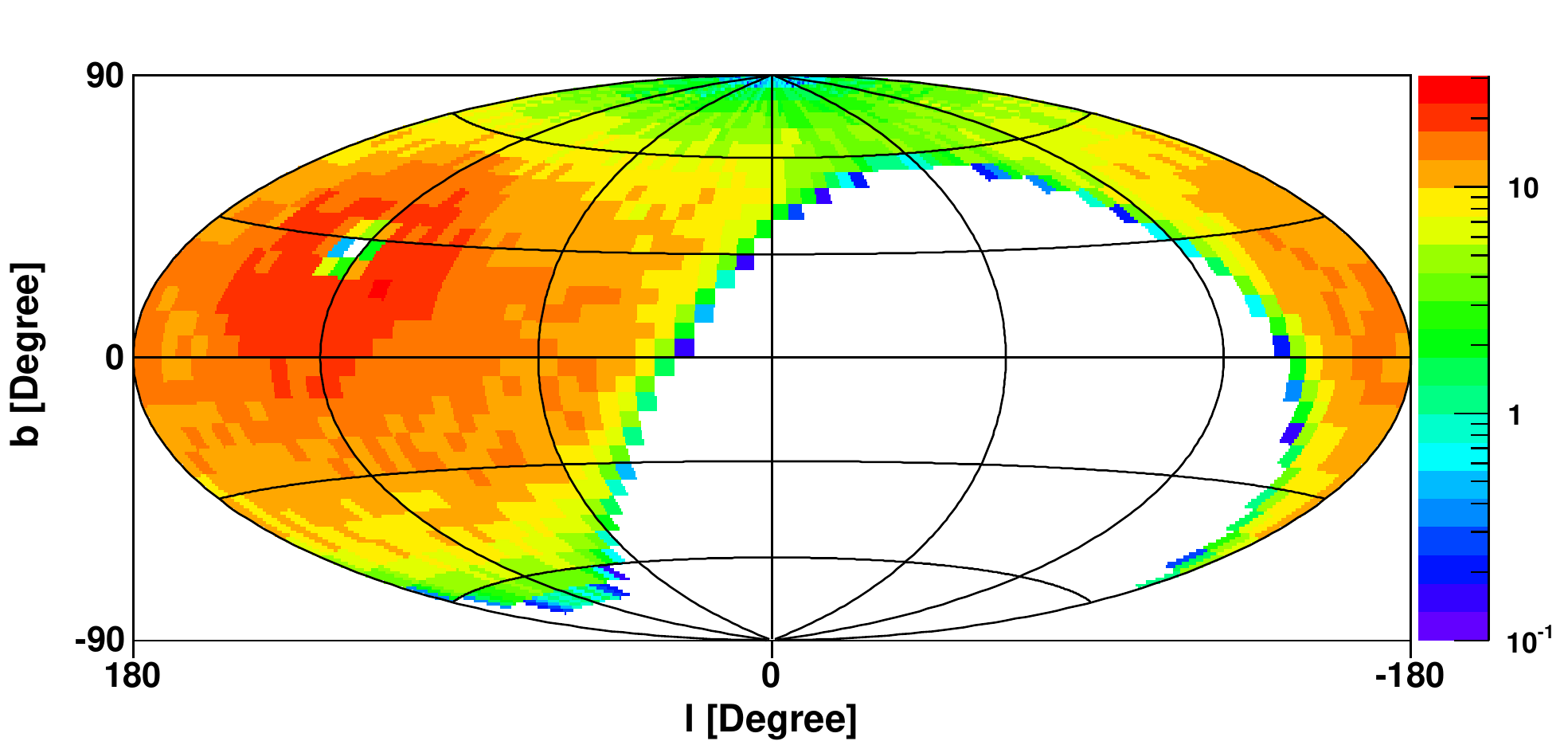}\label{figure:fit2d}}
  \caption{ Distributions of numbers of events in $b$ vs $l$ using
    5$^\circ$x5$^\circ$ bins on the sky \protect\subref{figure:data2d}
    for data and \protect\subref{figure:fit2d} for the fit to the
    model of $10^{18.0} - 10^{18.5}$ eV proton spectrum plus
    isotropy. The log-likelihood of the fit is 1574 over 1454 degrees
    of freedom, indicating that this is a good fit. The upper limit on
    the galactic proton flux found from this study is 0.9\%. }
  \label{figure:data_fit_2d}
\end{figure*}
In the second approach, we use the full 2-dimensional information of
the model of the galactic protons and fit the data distribution to a
function of the form, $N_{\mathrm{Data}}(l,b) = A
[F_{\mathrm{ISO}}(l,b) + \alpha F_{\mathrm{G.P.}}(l,b)]$, where
$N_{\mathrm{Data}}(l,b)$ is the number of data events in each bin in
$l$ and $b$, $F_{\mathrm{ISO}}(l,b)$ is the flux model in the case of
isotropy, $F_{\mathrm{G.P.}}(l,b)$ is the flux calculated using a
galactic proton model, $A$ is the overall normalization constant, and
$\alpha$ is the amplitude of the anisotropy due to the model, which is
related to the fraction $f$ of the galactic protons by
$f=\alpha/(1+\alpha)$, where $f \approx \alpha$ for $\alpha \ll 1$. If
we fit the data to the Galactic proton model that follows measured
cosmic ray spectrum (as shown in Figure~\ref{figure:data_fit_2d}), we
find the answer: $\alpha = -0.008 \pm 0.010$, which means that the
upper 95\% C.L. limit on $f$ is 0.9\%. If we repeat this search using
1 and 3 EeV galactic proton models, we find answers $\alpha = -0.010
\pm 0.009$ (0.5\% for 95\% C.L. on $f$) and $\alpha = -0.007 \pm
0.009$ (0.8\% for 95\% C.L. on $f$), respectively. Also, if we repeat
the same analysis in $10^{18.5}$ - $10^{19.0}$ eV energy range, the
answer for the fraction of the galactic protons is 0.7\%, at 95\%
confidence level.

\section{Summary}

We have searched for evidence that cosmic rays in the $10^{18.0}$ -
$10^{18.5}$ eV energy range, thought to be largely protons, are of
galactic origin.  Models of the galactic magnetic field predict that
if this were true anisotropy should be present in cosmic rays’ arrival
directions.  No anisotropy is seen. Examining the $l$ dependence of
the event distribution turns out to be the most sensitive test,
resulting in 0.3\% for the upper limit on the fraction of the galactic
protons, at 95\% confidence level. For the final answer, we choose
1.3\% (95\% C.L.), a more conservative answer obtained by comparing
the distributions of $b$ between the data and the isotropic Monte
Carlo.

\section{Acknowledgements}

The Telescope Array experiment is supported by the Japan Society for
the Promotion of Science through Grants-in-Aid for Scientific Research
on Specially Promoted Research (21000002) ``Extreme Phenomena in the
Universe Explored by Highest Energy Cosmic Rays'' and for Scientific
Research (19104006), and the Inter-University Research Program of the
Institute for Cosmic Ray Research; by the U.S. National Science
Foundation awards PHY-0307098, PHY-0601915, PHY-0649681, PHY-0703893,
PHY-0758342, PHY-0848320, PHY-1069280, PHY-1069286, PHY-1404495 and
PHY-1404502; by the National Research Foundation of Korea \linebreak
(2015R1A2A1A01006870, 2015R1A2A1A15055344, 2016R1A5A1013277,
\linebreak 2007-0093860, 2016R1A2B4014967); by the Russian Academy of
Sciences, RFBR grant 16-02-00962a (INR), IISN project No. 4.4502.13,
and Belgian Science Policy under IUAP VII/37 (ULB). The foundations of
Dr. Ezekiel R. and Edna Wattis Dumke, Willard L. Eccles, and George
S. and Dolores Dor\'e Eccles all helped with generous donations. The
State of Utah supported the project through its Economic Development
Board, and the University of Utah through the Office of the Vice
President for Research. The experimental site became available through
the cooperation of the Utah School and Institutional Trust Lands
Administration (SITLA), U.S. Bureau of Land Management (BLM), and the
U.S. Air Force. We appreciate the assistance of the State of Utah and
Fillmore offices of the BLM in crafting the Plan of Development for
the site.  We also wish to thank the people and the officials of
Millard County, Utah for their steadfast and warm support. We
gratefully acknowledge the contributions from the technical staffs of
our home institutions. An allocation of computer time from the Center
for High Performance Computing at the University of Utah is gratefully
acknowledged.


\newpage
\bibliographystyle{elsarticle-num}
\bibliography{galprot}

\end{document}